\documentclass[conference]{IEEEtran}
\IEEEoverridecommandlockouts
\usepackage{amsmath}
\usepackage{array}
\usepackage{booktabs}
\usepackage{cite}
\usepackage[T1]{fontenc}
\usepackage{url}
\title{Synchronized Three-Dimensional Vocal-Tract Motion for Speech Synchronization via Joint-Embedding Predictive Architecture Alignment}

\author{\IEEEauthorblockN{Sheng Li and Takahiro Shinozaki}
\IEEEauthorblockA{Institute of Science Tokyo\\
Email: sheng.li@ieee.org, shinot@ict.e.titech.ac.jp}}

\begin{document}
\maketitle

\begin{abstract}
Modern neural speech systems can generate intelligible waveforms, but they usually hide the physical speech-production state that produced the sound. Conversely, biomechanical vocal-tract models expose articulatory structure, contact behavior, airflow routing, and geometric constraints, but direct physical waveform synthesis remains less robust than modern neural vocoders. A duration-preserving acoustic carrier supplies the listening waveform, while a corrected three-dimensional vocal-tract model supplies synchronized jaw, lip, tongue, velum, laryngeal, oral-airflow, and nasal-airflow motion. A joint-embedding predictive architecture (JEPA)-style representation and a reinforcement learning/cross-entropy method (RL/CEM) trajectory-selection loop align articulatory actions to the acoustic carrier and to physical-plausibility constraints. The evaluation contains 12 3D recordings covering 24 minimal-pair stimuli. On the 24-word set, the carrier obtains good automatic speech recognition (ASR) results (an 8.33\% WER, a 4.17\% CER), a UTMOS score of 3.174, a mean JEPA score of 0.864, and a mean timbre-guard score of 0.947.
\end{abstract}


\section{Introduction}
Speech is a biomechanical control process as well as an acoustic signal. A single word is realized through coordinated motion of the jaw, lips, tongue body, tongue tip, velum, laryngeal structures, glottis, oral cavity, nasal cavity, and respiratory-aerodynamic system. In many modern speech pipelines, this state is hidden. Neural text-to-speech (TTS) and neural vocoder systems can produce highly intelligible audio, but the resulting waveform does not generally reveal where the tongue contacted the palate, whether the velum was lowered, how the lips opened, or whether a visible motion sequence remains physiologically plausible.

The opposite limitation appears in explicit articulatory simulation. A three-dimensional (3-D) vocal-tract model can expose the geometry, collision state, contact behavior, and airflow proxies that speech scientists, clinical tools, pronunciation-training systems, and anatomically grounded animation systems need. Yet fully physical waveform synthesis remains difficult because high-quality speech depends on tightly coupled source excitation, turbulence, wall losses, radiation, nasal coupling, and tissue-fluid interactions. At present, a system that insists on generating the final listening waveform only through a complete physical solver risks sacrificing intelligibility and usability.

The proposed system\footnote{https://halspeech.github.io/language} uses a strong acoustic carrier as the listening signal and a corrected 3-D vocal-tract model as the source of synchronized, interpretable motion. The carrier provides robust word-level intelligibility, while the model provides inspectable articulatory trajectories, transparent internal geometry, contact cues, oral and nasal airflow markers, and physiological consistency checks. The main contributions are: (1) a formulation of carrier speech and 3-D articulatory motion simulation; (2) a duration-preserving carrier coupled to a corrected articulatory model; (3) a joint-embedding predictive architecture/reinforcement learning (JEPA/RL) alignment objective combining automatic speech recognition (ASR) content, latent agreement, UTMOS quality, and timbre constraints; and (4) a recorded minimal-pair protocol using actual articulatory simulation.

\section{Related Works}
Articulatory and biomechanical speech synthesis has a long history in area-function modeling, 3-D vocal-tract geometry, biomechanical tissue simulation, and source-filter acoustics. 3D articulatory model provides a modeling environment for rigid bodies, finite elements, muscles, contact, and upper-airway simulation \cite{fels2006artisynth,gick2013modularizing}. VocalTractLab and related systems show the value of interpretable geometry and acoustic tube models, while also illustrating the difficulty of matching modern neural vocoder naturalness \cite{birkholz2006construction,birkholz2007turbulence}.

Measured articulatory corpora provide another route to interpretable speech generation. Electromagnetic articulography (EMA), real-time magnetic resonance imaging (MRI), and datasets such as Multi-CHannel Articulatory-TIMIT (MOCHA-TIMIT) and University of Southern California-TIMIT (USC-TIMIT) have supported articulatory-to-acoustic mapping, MRI-to-speech synthesis, and compact articulatory representations \cite{wrench1999mocha,narayanan2014database}. Recent neural systems map MRI, EMA, or multimodal articulatory representations into acoustic features or waveforms using learned decoders and vocoders \cite{otani2023speech,wu2023deep,liu2024fast}. Neural vocoders such as WaveNet and high-fidelity generative adversarial network (HiFi-GAN) show that high-quality waveform generation is practical when suitable acoustic conditioning is available \cite{oord2016wavenet,kong2020hifigan}. We therefore treat high-quality audio as a carrier and ask whether a physical vocal-tract model can produce synchronized, inspectable motion for that carrier while preserving timing, content, and physiological plausibility. Objective evaluation uses ASR-derived word and character error rates together with UTMOS-style automatic speech-quality estimates \cite{radford2022whisper,saeki2022utmos}.

\section{Formulation and System Architecture}
Let $x$ denote input text, $p$ a phone sequence, and $d$ the target phone or word durations. The controller produces an articulatory action sequence
\begin{equation}
    a_{1:T}=\pi_{\theta}(x,p,d),
\end{equation}
where the action vector contains interpretable controls for articulators and visualization channels. The articulatory model maps the sequence to a 3-D state trajectory,
\begin{equation}
    s_{t+1}=F_{\phi}(s_t,a_t,c_t),
\end{equation}
where $c_t$ includes timing, smoothness, contact, visibility, and geometry constraints. The delivered waveform is a carrier
\begin{equation}
    y^C=V(x,d),
\end{equation}
where $V$ is a reference, neural TTS, or vocoder source whose duration is preserved. A physical acoustic branch is retained as a diagnostic path,
\begin{equation}
    y^{\mathrm{phys}}=H(G(s_{1:T}),f_0,e),
\end{equation}
where $G$ extracts airway, source, and airflow proxies and $f_0$ denotes fundamental frequency. In the present implementation, $y^{\mathrm{phys}}$ is not the delivered listening signal.

The pipeline contains five functional blocks: text and duration preparation, articulatory action generation, articulatory model simulation, carrier waveform preparation, and recorded audiovisual preparation. The input word is converted into phones and a duration schedule. Word duration is preserved after timbre modification to prevent drift between visual motion and carrier audio. The controller emits time-indexed actions for the jaw, lips, tongue, velum, laryngeal region, airflow markers, and auxiliary visualization channels. These controls are mapped into anatomy-relevant state variables rather than treated as black-box animation parameters, so minimal-pair contrasts can be inspected in the same geometry used for physical checks.

The 3D model exposes the skin, jaw, skull, teeth, tongue, oral cavity, pharyngeal tract, laryngeal structures, vocal-fold region, soft palate, uvula, epiglottis, nasal tract, and airflow proxy markers. The transparent internal-inspection view is used for recorded videos because it makes tongue motion, contact regions, pharyngeal geometry, and nasal airflow visible in a single frame. Recordings are captured from running the articulatory movements in the Open Graphics Library (OpenGL) viewer, cropped to the viewer window, overlaid with burned-in word subtitles, and muxed with the carrier audio. This ensures that the recorded evidence includes the same camera, transparency, subtitle timing, and audio synchronization seen during interactive inspection.

\section{JEPA-Guided RL/CEM Alignment}
The JEPA module provides dense predictive feedback when a fully reliable physical waveform objective is unavailable. An encoder maps observations to latent targets, $z_t=E_{\omega}(o_t)$, where $o_t$ may include actions, mesh state, tube or airway proxies, airflow markers, contact variables, and carrier-acoustic features. A predictor estimates future latent states from the current latent context and candidate actions,
\begin{equation}
    \hat{z}_{t+k}=P_{\psi}(z_{1:t},a_{t:t+k}).
\end{equation}
The JEPA loss is
\begin{equation}
    \mathcal{L}_{\mathrm{JEPA}}=\sum_{t,k}\left\|\hat{z}_{t+k}-\mathrm{sg}(z_{t+k})\right\|_1,
\end{equation}
where $\mathrm{sg}(\cdot)$ denotes a stop-gradient target. In this implementation, the loss provides a compact alignment signal between the planned action trajectory, the physical state proxies, and the acoustic carrier representation.

Candidate action trajectories are selected with a multi-objective RL/CEM reward:
\begin{equation}
    R=0.64R_{\mathrm{ASR}}+0.16R_{\mathrm{JEPA}}+0.12R_{\mathrm{UTMOS}}+0.08R_{\mathrm{timbre}}.
\end{equation}
The ASR term rewards recovery of the intended word from the carrier; the JEPA term rewards latent agreement between action-conditioned prediction and target observations; the UTMOS term discourages audio-quality regressions; and the timbre term discourages drift away from the experiment's selected carrier setting. Optimizing ASR alone could accept a carrier with poor audiovisual alignment, while optimizing visual smoothness alone could hide articulatory contrasts. The reward therefore treats an improvement as valid only when carrier, trajectory, and plausibility terms remain jointly acceptable.

Implementation focuses on carrier stability, airflow-marker placement, and recorded evidence. The nasal airway and epiglottis positions were checked against current model geometry. The nasal path was shortened and lowered so that nasal airflow exits near the nose rather than appearing near the eye region. Airflow markers are derived from current oral and nasal mesh bounds rather than fixed historical coordinates. For consonantal minimal pairs, the view supports inspection of bilabial closure, labiodental constriction, alveolar contact, velar approach, nasal airflow routing, and liquid or glide postures. For vowel contrasts, it supports inspection of approximate tongue-body, lip-rounding, and tract-shape changes.

\section{Experimental settings and Results}
The evaluation is diagnostic rather than corpus-scale. It uses 24 words grouped into 12 minimal pairs, covering stop voicing at bilabial, alveolar, and velar places; fricative and labiodental voicing; nasal place; liquid contrast; high, front, low/back, and rounded vowel contrasts; and glide contrast. Audio intelligibility is measured with ASR-derived exact-word recovery, WER, and CER. Audio quality is measured with UTMOS as an automatic regression signal. Timbre consistency is measured with mean $f_0$, spectral centroid before and after the timbre operation, and an experiment-specific timbre-guard score. Alignment quality is summarized with the JEPA score and the combined reward. Recorded motion is evaluated by asking whether the minimal-pair contrast is visible, whether lip and tongue contacts are plausible, whether nasal airflow follows the nasal tract, and whether motion remains synchronized with the word audio.

\begin{table}[!t]
\caption{Objective metrics over the 24 minimal-pair stimuli.}
\centering
\scriptsize
\setlength{\tabcolsep}{3pt}
\begin{tabular}{@{}p{0.65\columnwidth}p{0.25\columnwidth}@{}}
\toprule
Metric & Value \\
\midrule
Stimuli & 24 words / 12 pairs \\
Exact ASR outputs & 22 / 24 \\
Mean WER / CER & 8.33\% / 4.17\% \\
Mean UTMOS & 3.174 \\
Mean $f_0$ before / after & 119.54 / 119.06 Hz \\
Spectral centroid before / after & 1883.85 / 1717.43 Hz \\
Timbre-guard score & 0.947 \\
Mean JEPA score & 0.864 \\
Mean combined reward & 0.883 \\
\bottomrule
\end{tabular}
\label{tab:metrics}
\vspace{-1mm}
\end{table}

\begin{table}[!t]
\caption{Minimal-pair audiovisual inspection set.}
\centering
\begin{tabular}{@{}c l l c l l@{}}
\toprule
No. & Contrast & Stimuli & No. & Contrast & Stimuli \\
\midrule
1 & Bilabial & pat/bat & 7 & Liquid & light/right \\
2 & Alveolar & ten/den & 8 & High vowel & beat/bit \\
3 & Velar & cap/gap & 9 & Front vowel & bad/bed \\
4 & Fricative & sip/zip & 10 & Low/back & cot/caught \\
5 & Labiodental & fan/van & 11 & Rounded & pool/pull \\
6 & Nasal place & map/nap & 12 & Glide & woo/you \\
\bottomrule
\end{tabular}
\label{tab:minpairs}
\\(https://halspeech.github.io/language)
\end{table}

Table~\ref{tab:metrics} summarizes the objective metrics. The carrier obtains exact ASR output for 22 of 24 words, with mean WER of 8.33\% and mean CER of 4.17\%. These values indicate that the listening signal is generally recoverable by an automatic recognizer over this diagnostic set, but the two non-exact cases motivate larger-scale error analysis before corpus-level claims. The mean UTMOS score is 3.174; this should be read as an automatic quality estimate, not a human MOS result. The timbre operation reduces the mean spectral centroid from 1883.85 Hz to 1717.43 Hz, an 8.83\% reduction, while mean $f_0$ changes only from 119.54 Hz to 119.06 Hz. This supports the intended lower-brightness carrier without meaningful $f_0$ or duration disruption. The mean JEPA score is 0.864 and the mean combined reward is 0.883. Because the reward combines ASR, JEPA, UTMOS, and timbre terms, the combined value is best interpreted as a configuration-selection and regression-monitoring statistic.

The 12 minimal pairs in Table~\ref{tab:minpairs} provide an audit trail for audiovisual inspection. Each recording is cropped 3D articulatory movement viewer footage with burned-in word subtitles and muxed carrier audio from the same experimental setting, enabling expert review of bilabial closure, labiodental constriction, alveolar contact, velar approach, nasal routing, liquid posture, and vowel-related jaw or lip changes. Subjective evaluation for minimal pairs is ongoing.

\section{Discussion}
The results show the system provides intelligible carrier audio, synchronized 3-D articulatory motion, visual airflow proxies, and a reproducible minimal-pair inspection protocol for workflows that need an explicit speech-production state while retaining an intelligible listening signal.

The limitations are also clear. The delivered waveform is a carrier, not the output of a complete articulatory acoustic solver based on finite-element-method/computational-fluid-dynamics (FEM/CFD) side information. JEPA is an alignment and representation mechanism, not proof of full physical understanding. The 24-word minimal-pair set is diagnostic; it should be followed by sentence-level, multi-speaker, and corpus-scale evaluations. UTMOS and ASR are automatic proxies and should be supplemented with human listening tests, expert articulatory ratings, audiovisual synchrony judgments, and more ablations over the JEPA term, timbre guard, and physical-plausibility constraints. These will be our future focus.

\section{Conclusion}
In conclusion, the paper contributes a synchronized audio-motion demonstration and evaluation protocol with corrected articulatory anatomy, JEPA/RL reward accounting, 12 GUI recordings, and 24 minimal-pair words. Across the reported set, it obtains good ASR results (8.33\% WER, 4.17\% CER), 3.174 mean UTMOS, 0.864 mean JEPA score, and 0.947 mean timbre-guard score.

\section*{Acknowledgment and Artificial Intelligence (AI) Disclosure}
OpenAI ChatGPT and Grammarly were used to polish drafts and assist with table organization in this manuscript. The idea, experimental design, results, claims, citations, and final text were run and written by the authors.

\bibliographystyle{IEEEtran}
\bibliography{references}

\end{document}